\documentstyle[12pt]{article}

\begin{document}
\hbadness=10000
\hbadness=10000
\begin{titlepage}
\nopagebreak
\begin{flushright}
{\normalsize
~
~
}
\end{flushright}
\vspace{0.7cm}
\begin{center}

{\large \bf Symmetry Reduction, Gauge Transformation\\
and Orbifold}

\vspace{1.2cm}

Tetsuaki {\sc Kawamoto}
and
Yoshiharu {\sc Kawamura}\footnote{E-mail:
haru@azusa.shinshu-u.ac.jp}

\vspace{0.9cm}
Department of Physics, Shinshu University,
Matsumoto 390-8621, Japan

\end{center}
\vspace{0.9cm}

\nopagebreak

\begin{abstract}
We study a mechanism of symmetry reduction in a higher-dimensional
field theory upon orbifold compactification.
Split multiplets appear unless all components in a multiplet 
of a symmetry group have a common parity on an orbifold.
A gauge transformation property is also examined.
\end{abstract}
\vfill
\end{titlepage}
\pagestyle{plain}
\newpage
\def\thefootnote{\fnsymbol{footnote}}

\section{Introduction}

Recently, a new possibility \cite{K2} has been proposed to reconcile the
coupling unification
scenario with the triplet-doublet mass splitting
based on a 5-dimensional (5D) supersymmetric (SUSY) model with $SU(5)$
gauge symmetry.
The minimal supersymmetric standard model (MSSM)
is derived on a 4D wall 
through compactification on $S^1/(Z_2 \times Z_2')$.\footnote{
Recently, Barbieri, Hall and Nomura have
constructed a constrained standard model 
upon a compactification of a 5D SUSY model on 
the orbifold $S^1/(Z_2 \times Z_2')$.\cite{BHN}
They used $Z_2 \times Z_2'$ parity to reduce SUSY.
There are also several works on model building 
through a reduction of SUSY \cite{orb,AMQ,HW,MP,PQ} 
by the use of a discrete symmetry and a reduction of gauge symmetry \cite{K} 
by the use of $Z_2$ parity.
Attempts to construct unified models have been made through
dimensional reduction over coset space.\cite{coset}
The study of higher-dimensional SUSY grand unified theories traces back 
to the work by Fayet.\cite{Fayet}}
The excellent characteristics of this model have been studied.
\cite{AF,HN}\footnote{
There are several works on the other type of 5D unifed models with 1D
orbifold, i.e.,
5D $SU(5)$ model with $S^1/Z_2$ \cite{K}, 
5D $SU(5)$ model with $S^1/(Z_2 \times Z_2')$ \cite{K3}
and 5D SUSY $SU(5)$ model with $S^1/Z_2$ \cite{Ko}.}
The key features are as follows.
\begin{itemize}
\item Unless components in a multiplet have a common $Z_2 \times Z_2'$ parity 
on the orbifold, the lowest modes in 4D fields do not form full 
multiplets of $SU(5)$.
It realizes a triplet-doublet splitting and an SM-$X$,$Y$ gauge multiplets
splitting with a suitable assignment of $Z_2 \times Z_2'$ parity.

\item A specific type of $SU(5)$ gauge symmetry exists
on one of 4D walls (a visible wall) as well as in the bulk.
It leads to a coupling unification at the zero-th order approximation.

\item 5D bulk fields and 4D fields on the visible wall belong to some
representations of $SU(5)$.
It guarantees the quantization of charge.
\end{itemize}
We expect that similar features hold in a class of higher-dimensional
grand unified theory (GUT) as suggested in Ref. \cite{HN}. 
Concretely,
\begin{enumerate}
\item Unless all components in a multiplet of some unified gauge group $G_U$ 
have a common parity on an orbifold, 
split multiplets appear after the integration of the extra
space because the lowest modes, in general, do not form full 
multiplets of $G_U$.

\item The higher-dimensional gauge symmetry is realized as an invariance
under the gauge transformation whose gauge functions have a definite parity
on an orbifold, and hence the gauge symmetry at some points on the orbifold
turns out to be a reduced one whose generators are commutable to a parity
operator.
\end{enumerate}

In this paper, we study the above features in GUTs on an orbifold, which 
would be important for a  construction of a realistic model and an exploration
of the origin of symmetries in the SM.

This paper is organized as follows. 
In the next section, we study a mechanism of symmetry reduction
due to an intrinsic parity on an orbifold.
We discuss the reduction of gauge symmetry,
a gauge transformation property
and its phenomenological implications in $\S$3. 
Section 4 is devoted to conclusions and discussion.

\section{Splitting from $Z_N$ parity}

The space-time is assumed to be factorized into a product of 
4D Minkowski space-time $M^4$ and the $2n$-dimensional ($2n$-D)
orbifold $O^{2n} \equiv T^{2n}/\prod_N Z_N$,
whose coordinates are denoted
by $x^\mu$ ($\mu = 0,1,2,3$) and $y^{\hat{\mu}}$  ($\hat{\mu} = 1, 2,
\cdots ,2n$), respectively.
The notation $x^M$ ($M = 0,1,2,3,5, \cdots,2n+4$) is also used for coordinates.
The orbifold $O^{2n}$ is obtained by dividing a $2n$-D torus $T^{2n}$
with $Z_N$ rotations which are automorphisms of $T^{2n}$. \cite{orb}
The $Z_N$ rotation is diagonalizable under a suitable complex basis 
$(z^i, \bar{z}^i)$ ($i = 1,2,\cdots, n$) for the extra space and is given by 
the transformation $z^i \to z^{'i} = \theta^i_j z^j$.
Here $\theta^i_j$ is an element of $Z_N$ transformation written by
\begin{eqnarray}
 \theta^i_j &=& \mbox{diag}\left(\exp{2\pi i m_1 \over N}, \exp{2\pi i m_2
\over N}, \cdots, \exp{2\pi i m_n \over N}\right) \nonumber \\
 &\equiv & \mbox{diag}( \theta_1, \theta_2, \cdots, \theta_n )
\label{theta}
\end{eqnarray}
where $m_i$ are integers.
The $T^{2n}$ is regarded as a $2n$-D lattice
that the point $z^i$ is identified with $z^i + n^I e_I^i$
where $n^I$ are integers and $e_I^i$ are shift vectors on the lattice.
There are points fixed by the discrete transformation.
They are called fixed points,\footnote{
Since fixed points are singular points on the space, orbifolds are not 
manifolds.
We assume that this singularity does not cause any trouble
in an underlying theory.}
which are denoted by
$z_{fp}^i$ and satisfy the relation $z_{{fp}}^i = \theta^i_j z_{{fp}}^j +
n^I e_I^i$.

Here we study a field theory on 2D $Z_3$ orbifold as an example.
The $Z_3$ orbifold is obtained by dividing the $SU(3)$ root lattice
$\Gamma_{SU(3)}$
with a $Z_3$ rotation
whose element is $\theta = \exp{2\pi i \over 3}$.
The shift vectors on $\Gamma_{SU(3)}$ are given by $1$ 
and $\omega \equiv \exp{2\pi i \over 3}$.\footnote{
As we take a normalization where a size of extra space equals that of
$\Gamma_{SU(3)}$,
we should consider that the compact space has a physical size $2\pi R$ 
on the estimation of a magnitude of physical quantities.}
Hence the following identification holds on the orbifold,
\begin{eqnarray}
z \sim  z + 1 \sim z + \omega \sim \omega z .
\label{Z3}
\end{eqnarray}
There are three kinds of fixed points,
\begin{eqnarray}
z_{{fp}} = 0, {2+ \omega \over 3}, {1 + 2 \omega \over 3}  .
\label{fpZ3}
\end{eqnarray}
An intrinsic $Z_3$ parity of the bulk field $\phi(x^\mu, z, \bar{z})$
is defined by the transformation
\begin{eqnarray}
  \phi(x^\mu, z, \bar{z}) \to \phi(x^\mu, \omega z, \omega^2 \bar{z}) 
  = P \phi(x^\mu,  z, \bar{z}) .
\label{P-tr}
\end{eqnarray}
By definition, $P$ possesses only the eigenvalues 1, $\omega$ or $\omega^2$.
We denote the fields that are eigenfunctions of $P$ as 
$\phi_{\omega^0}$, $\phi_{\omega^1}$, $\phi_{\omega^2}$ where the 
subscript corresponds to the eigenvalue of $P$.
The 6D fields $\phi_{\omega^l}$ $(l=0,1,2)$
are Fourier expanded as
\begin{eqnarray}
 \phi_{\omega^l} (x^\mu, z, \bar{z}) = 
      \sum_{n,m} \phi^{(nm)}_{\omega^l}(x^\mu) f_{nm}^{\omega^l}(z, \bar{z})  
\label{phi(l)exp}
\end{eqnarray}
where $n$ and $m$ are integers, and $f_{nm}^{\omega^l}(z, \bar{z})$ are 
eigenfunctions of $P$ whose eigenvalues are $\omega^l$.
The $f_{nm}^{\omega^l}(z, \bar{z})$ are written as
\begin{eqnarray}
&~& f_{nm}^{\omega^0}(z, \bar{z}) = f_{nm}(z, \bar{z}) 
+ f_{nm}(\omega z, \omega^2 \bar{z}) + f_{nm}(\omega^2 z, \omega \bar{z}) ,
\label{fnm(0)}\\
&~& f_{nm}^{\omega^1}(z, \bar{z}) = f_{nm}(z, \bar{z}) 
+ \omega^2 f_{nm}(\omega z, \omega^2 \bar{z}) + \omega f_{nm}(\omega^2 z,
\omega \bar{z}) ,
\label{fnm(1)}\\
&~& f_{nm}^{\omega^2}(z, \bar{z}) = f_{nm}(z, \bar{z}) 
+ \omega f_{nm}(\omega z, \omega^2 \bar{z}) + \omega^2 f_{nm}(\omega^2 z,
\omega \bar{z}) .
\label{fnm(2)}
\end{eqnarray}
by the use of a function $f_{nm}(z, \bar{z})$ which satisfies periodic
boundary conditions
\begin{eqnarray}
 f_{nm}(z, \bar{z}) =  f_{nm}(z+1, \bar{z}+1) = f_{nm}(z+\omega,
\bar{z}+\omega^2)  .
\label{PBC}
\end{eqnarray}
The explicit form of $f_{nm}(z, \bar{z})$ is given by
\begin{eqnarray}
 f_{nm}(z, \bar{z})  = \exp\left(\pi i\left(\left(n - {n + 2m \over
\sqrt{3}}i\right)z + \left(n + {n + 2m \over \sqrt{3}}i\right)\bar{z}
\right)\right) .
\label{fnm}
\end{eqnarray} From the expressions (\ref{phi(l)exp})$-$(\ref{fnm}),
we find the following features of eigenfunctions.
\begin{itemize}
\item The 4D fields $\phi^{(nm)}_{\omega^l}(x^\mu)$ 
acquire mass $(n^2+{(n + 2m)^2 \over 3})^{1/2}/R$ upon compactification. 

\item The 4D fields with $n=m=0$ (4D zero modes) appear from 6D fields
whose $Z_3$ parity is 1, i.e., the $\phi_{\omega^0}(x^\mu, z, \bar{z})$ has
4D zero mode.

\item The 6D fields whose $Z_3$ parity is $\omega$ or $\omega^2$ vanish on
the fixed points, i.e.,
$\phi_{\omega^1}(x^\mu, z_{fp}, \bar{z}_{fp})$ $= \phi_{\omega^2}(x^\mu,
z_{fp}, \bar{z}_{fp}) = 0$.
\end{itemize}
Let us study the case in which a field $\Phi(x^\mu, z, \bar{z})$ is an
$N_f$-plet
under some symmetry group.
The components of $\Phi$ are denoted by $\Phi = (\phi_1, \phi_2, ...,
\phi_{N_f})^T$.
The $Z_3$ transformation of $\Phi$ is given by the same form as 
(\ref{P-tr}), but in this case $P$ is an $N_f \times N_f$ matrix
which satisfies $P^3 = I$, where $I$ is the unit matrix.
The $Z_3$ invariance of the Lagrangian density
does not necessarily require that $P$ be proportional to $I$.
Unless all components of $\Phi$ have a common $Z_3$ parity, 
the splitting in a multiplet occurs upon compactification
because of the lack of zero modes in components with $Z_3$ parity other
than one.

The generalization on a model with a generic orbifold is straightforward.
Hence, {\it{in a class of higher-dimensional GUT on an orbifold,
unless all components in a multiplet of some unified gauge group $G_U$ 
have a common parity on an orbifold,
split multiplets appear after the integration of the extra
space because zero modes, in general, do not form full 
multiplets of $G_U$.}}

\section{Gauge transformation property}

We apply the mechanism of symmetry reduction discussed in the previous section
to GUTs on $M^4 \times O^{2n}$.
Here we consider a non-SUSY model for simplicity.
The SUSY extension is straightforward.
We take two basic assumptions.
One is that the gauge boson $A_M(x^\mu, z^i, \bar{z}^i) 
= A_M^{\alpha}(x^\mu, z^i, \bar{z}^i) T^{\alpha}$ 
and a scalar field $\Phi(x^\mu, z^i, \bar{z}^i)$ exist in the bulk.
Here the $T^{\alpha}$ are gauge generators
and the $\Phi(x^\mu, z^i, \bar{z}^i)$ belongs to a vector representation of
a unified group $G_U$.
The other is that our visible world is one of 4D walls at a certain
point on the orbifold and matter fields are located on the wall.

The action integral is given by
\begin{eqnarray}
 S &=& \int  {\cal L}_{{bulk}} d^{4+2n}x + \sum_{p} \int {\cal
L}_{{fp}}^{(p)} d^{4+2n}x , \\
 {\cal L}_{{bulk}} &\equiv&  - {1 \over 2} {\rm tr} F_{MN}F^{MN} + 
|D_M \Phi|^2 - V(|\Phi|^2) 
\end{eqnarray}
where $D_M \equiv \partial_M - i g_U A_M(x^M)$,
$g_U$ is a $(4+2n)$-D gauge coupling constant and ${\cal L}_{fp}^{(p)}$ is
a contribution
from the $p$-th 4D wall.
The above Lagrangian density ${\cal L}_{bulk}$ is invariant 
under a $Z_N$ transformation and a gauge transformation
defined as follows.
The $Z_N$ transformation for $A_M$ and $\Phi$
is given by
\begin{eqnarray}
 &~& A_{\mu}(x^\mu, z^i, \bar{z}^i) \to A_{\mu}(x^\mu, z^{'i}, \bar{z}^{'i}) = 
P A_{\mu}(x^\mu, z^i, \bar{z}^i) P^{-1} , \nonumber \\
 &~& A_{z^l}(x^\mu, z^i, \bar{z}^i) \to A_{z^l}(x^\mu, z^{'i}, \bar{z}^{'i}) = 
\theta_l^{-1} P A_{z^l}(x^\mu, z^i, \bar{z}^i) P^{-1} , \nonumber \\
&~& A_{\bar{z}^l}(x^\mu, z^i, \bar{z}^i) \to A_{\bar{z}^l}(x^\mu, z^{'i},
\bar{z}^{'i}) = 
\theta_l P A_{\bar{z}^l}(x^\mu, z^i, \bar{z}^i) P^{-1} , \nonumber \\
 &~& \Phi(x^\mu, z^i, \bar{z}^i) \to \Phi(x^\mu, z^{'i}, \bar{z}^{'i}) = P
\Phi(x^\mu, z^i, \bar{z}^i)  
\label{P-tr2}
\end{eqnarray}
where $P$ is $Z_N$ parity operator,
$z^{'i} = \theta_i z^i$ and $\bar{z}^{'i} = \bar{\theta}_i \bar{z}^i$.
The gauge transformation for $A_M$ and $\Phi$
is given by
\begin{eqnarray}
 &~& A_{M}(x^\mu, z^i, \bar{z}^i) \to A'_{M}(x^\mu, z^i, \bar{z}^i) = 
U A_{M}(x^\mu, z^i, \bar{z}^i) U^{-1} + {i \over g_U} U\partial_M U^{-1},
\nonumber \\
 &~& \Phi(x^\mu, z^i, \bar{z}^i) \to \Phi'(x^\mu, z^i, \bar{z}^i) = U
\Phi(x^\mu, z^i, \bar{z}^i)  
\label{gauge-tr}
\end{eqnarray}
where $U$ is a space-time dependent gauge transformation matrix.
The $Z_N$ transformation is, in general, not commutable to a gauge
transformation
with generic gauge functions,  
unless $P$ is proportional to the unit matrix.
But, when there is a relation
$P T^{\alpha} P^{-1} = \theta^{k_{\alpha}} T^{\alpha}$ 
and the group structure constants $f^{\alpha\beta\gamma}$ vanish
for $k_\alpha + k_\beta \neq k_\gamma$ (mod $N$),
there survives a specific type of unified gauge symmetry, which is
compatible with the $Z_N$
transformation, based on a gauge transformation matrix given by
\begin{eqnarray}
U(x^M) = \exp(i \xi^{\alpha}_{\theta^{k_{\alpha}}}(x^M) T^{\alpha}) 
\label{hatU}
\end{eqnarray}
where gauge functions $\xi^{\alpha}_{\theta^{k_{\alpha}}}(x^M)$ are
eigenfunctions
with eigenvalue $\theta^{k_{\alpha}}$ for $Z_N$ parity.
Actually the gauge transformation matrix (\ref{hatU}) is obtained
from the requirement that a $Z_N$ parity of $A_M^{'\alpha}$ and $\Phi'_{k}$
equals 
that of $A_M^{\alpha}$ and $\Phi_{k}$ or
that the $Z_N$ parity assignment of each component in a multiplet 
is preserved after the gauge transformation, i.e.,
\begin{eqnarray}
PU(x^\mu, z^i, \bar{z}^i) = U(x^\mu, z^{'i}, \bar{z}^{'i}) P .
\label{P-U}
\end{eqnarray}
The reduction of gauge symmetry occurs at a fixed point $z_{fp}^i$ because
the $\xi^{\alpha}_{\theta^{k_{\alpha}}}(x^M)$ vanish at $z_{fp}^i$ for
$k_\alpha \neq 0$
(mod $N$).
The residual gauge group is a subgroup of $G_U$, whose generators are commutable
to $Z_N$ parity operator.
The interaction on $z_{fp}^i$ is constrained from the symmetry there.
For example, the Lagrangian density on $z_{fp}^i$ should be invariant under
both $Z_N$ parity and
the residual gauge transformation. 

The above feature can be generalized in the case with a generic orbifold as
a statement that
{\it{in higher-dimensional space-time,
there exists a specific type of unified gauge symmetry 
based on gauge functions with a definite parity on an orbifold, 
Hence the gauge symmetry is reduced to
a smaller one whose generators are commutable 
to a parity operator at some points on the orbifold
because some of gauge functions vanish there.}}

Finally we discuss 4D particle spectrum of a model with $G_U = SU(5)$ 
on $Z_N$ orbifold.
When we take 
$P = \mbox{diag}(\theta^k, \theta^k, \theta^k, 1, 1)$ for $k \neq 0$ (mod $N$),
the gauge symmetry is reduced to that of
the Standard Model, $G_{SM} \equiv SU(3) \times SU(2)
\times U(1)$, in 4D theory.\footnote{
Our symmetry reduction mechanism is different from a mechanism
due to the non-vanishing vacuum expectation value (VEV) of 4D scalar fields.
In our situation, the Hosotani mechanism \cite{H} does not work,
because $A_{z^l}^{a}(x^M)$ has a parity other than one,
and its VEV should vanish.}
This is because some of the gauge generators $T^{\alpha}$ 
$(\alpha = 1, 2, \cdots ,24)$ are not commutable with $P$,
\begin{eqnarray}
  P T^a P^{-1} = T^a , ~~
  P T^{\hat{a}_+} P^{-1} = \theta^k T^{\hat{a}_+} , ~~
  P T^{\hat{a}_-} P^{-1} = \theta^{-k} T^{\hat{a}_-} 
\end{eqnarray}
where the $T^a$ are gauge generators of $G_{SM}$
and the $T^{\hat{a}_{\pm}}$ are other gauge generators.
The $Z_N$ parity assignment of 4D fields is given in Table I.
\begin{table}[t]
\caption{4D fields and $Z_N$ Parity.}
\renewcommand{\arraystretch}{1.6}
\begin{center}
\begin{tabular}{l|l|l}
\hline\hline
4D fields & Quantum numbers & $Z_N$ parity  \\
\hline
$A_{\mu}^{a(\vec{n}\vec{m})} $ & $({\bf 8}, {\bf 1}) + ({\bf 1}, {\bf 3}) +
({\bf 1}, {\bf 1})$ & $1$  \\
$A_{\mu}^{\hat{a}_+(\vec{n}\vec{m})}$ &  $({\bf 3}, {\bf 2})$ & $\theta^k$  \\
$A_{\mu}^{\hat{a}_-(\vec{n}\vec{m})}$ &  $(\bar{\bf 3}, {\bf 2})$ &
$\theta^{-k}$  \\
\hline
$A_{z^l}^{a(\vec{n}\vec{m})}$ & $({\bf 8}, {\bf 1}) + ({\bf 1}, {\bf 3}) +
({\bf 1}, {\bf 1})$ & $\theta_l^{-1}$ \\
$A_{z^l}^{\hat{a}_+(\vec{n}\vec{m})}$ & $({\bf 3}, {\bf 2})$ & $\theta^{k}
\theta_l^{-1}$ \\
$A_{z^l}^{\hat{a}_-(\vec{n}\vec{m})}$ & $(\bar{\bf 3}, {\bf 2})$ &
$\theta^{-k} \theta_l^{-1}$ \\
\hline
$A_{\bar{z}^l}^{a(\vec{n}\vec{m})}$ & $({\bf 8}, {\bf 1}) + ({\bf 1}, {\bf
3}) + ({\bf 1}, {\bf 1})$ & $\theta_l$ \\
$A_{\bar{z}^l}^{\hat{a}_+(\vec{n}\vec{m})}$ & $({\bf 3}, {\bf 2})$ &
$\theta^{k}\theta_l$ \\
$A_{\bar{z}^l}^{\hat{a}_-(\vec{n}\vec{m})}$ & $(\bar{\bf 3}, {\bf 2})$ &
$\theta^{-k} \theta_l$ \\
\hline
$\phi_{C}^{(\vec{n}\vec{m})}$ & $({\bf 3}, {\bf 1})$ & $\theta^{k}$ \\
$\phi_{W}^{(\vec{n}\vec{m})}$ & $({\bf 1}, {\bf 2})$ & $1$ \\
\hline
\end{tabular}
\end{center}
\end{table}
The scalar field is divided into two pieces: 
$\Phi$ is divided into the colored triplet piece, $\phi_C$, 
and the $SU(2)$ doublet piece, $\phi_W$.
In the second column, we give the $SU(3) \times SU(2)$ quantum numbers
of the 4D fields.
In the third column, $Z_N$ parity of 4D fields is given.
We find that the 4D massless fields include  SM gauge bosons
$A_\mu^{a(\vec{0}\vec{0})}$ and
a weak Higgs doublet $\phi_W^{(\vec{0}\vec{0})}$ and that the
triplet-doublet mass splitting of 
the Higgs multiplets is realized
by projecting out zero modes of the colored components.
Whether or not extra massless particles appear depends on an assignment of
$Z_N$ parity.
Let us take 6D $SU(5)$ GUTs as an example.
In the case with $Z_3$ orbifold,
the $SU(5)$ is reduced to $G_{SM}$ in 4D theory
with $P = \mbox{diag}(\omega, \omega, \omega, 1, 1)$ and
the 4D massless fields consist of  SM gauge bosons $A_\mu^{a(00)}$,
SM weak Higgs doublet $\phi_W^{(00)}$ and extra 4D scalar fields
($A_z^{\hat{a}_+(00)}$,
$A_{\bar{z}}^{\hat{a}_-(00)}$).
In the case with $Z_4$ orbifold,
the $SU(5)$ is reduced to $G_{SM}$ in 4D theory
with $P = \mbox{diag}(i^k, i^k, i^k, 1, 1)$
for $k \neq 0$ (mod 4).
If we take $P = \mbox{diag}(i, i, i, 1, 1)$, extra 4D scalar fields appear.
But if we take $P = \mbox{diag}(-1, -1, -1, 1, 1)$,
the 4D massless fields consist of  SM gauge bosons $A_\mu^{a(00)}$ and
SM weak Higgs doublet $\phi_W^{(00)}$.
No extra 4D scalar fields appear.

\section{Conclusions and discussion}

We have studied a mechanism of symmetry reduction
due to an intrinsic parity on an orbifold.
In a class of higher-dimensional GUT, unless all components in a multiplet of 
some unified gauge group $G_U$ have a common parity on an orbifold, 
split multiplets appear after the integration of the extra
space because zero modes do not form full multiplets of $G_U$.
We have discussed the reduction of unified gauge symmetry,
gauge transformation property
and its phenomenological implications.
The higher-dimensional gauge symmetry is realized as an invariance
under the gauge transformation whose gauge functions have a definite parity
on an orbifold, and hence the gauge symmetry at some points
in the compact space turns out to be a reduced one 
whose generators are commutable to a parity operator.

The origin of a specific parity assignment is unknown, and
we believe that it will be explained in terms of
some yet to be constructed underlying theory.
The merit of this type of symmetry reduction is that there might be no
sizable contribution
to the vacuum energy upon compactification because there exists no field with
a non-vanishing VEV of $O(M_C)$ in our model.\footnote{
In the framework of supergravity theory, a large amount of (negative) vacuum energy
can be generated on the breakdown of a unified gauge symmetry by Higgs mechanism
through the non-vanishing VEV of the superpotential.}
Here $M_C$ is a compactification scale, which is related to a unification
scale $M_U$.

To construct a more realistic model, it is reasonable to require the
following conditions
on a 4D theory.
\begin{itemize}
\item The coupling unification holds at the zero-th order approximation.

\item The quantization of charge is derived.

\item The weak scale is stable against radiative corrections.
\end{itemize}
It is desirable that our 4D world is a specific point on an extra space 
where a unified gauge symmetry survives
from the first and second requirements.
The stability of the weak scale can be guaranteed by a SUSY extention of a
model.
However, in a higher-dimensional SUSY GUT, Higgs multiplet appears as a
hypermultiplet
and it is difficult to project out all zero modes of colored Higgs multiplets
by the use of a single parity.
Hence it would be quite interesting to study SUSY GUTs on a more complex
orbifold
constructed by dividing a torus with several discrete symmetries.

\end{document}